\documentclass[authoryear,preprint,review,12pt]{elsarticle}

\usepackage{amssymb}
\usepackage{amsmath}

\usepackage[dvipsnames]{xcolor}

\usepackage{ulem} 

\usepackage{hyperref}
\hypersetup{
    colorlinks=true,
    linkcolor=blue,         
    urlcolor=cyan,          
}

        \makeatletter
        \def\ps@pprintTitle{%
         \let\@oddhead\@empty
         \let\@evenhead\@empty
         \def\@oddfoot{}%
         \let\@evenfoot\@oddfoot}
        \makeatother

\usepackage{xcolor}
\newcounter{aqctr}
\newenvironment{author-query}
{\refstepcounter{aqctr}\par\vspace{\baselineskip}\noindent
\color{red}\textbf{Author Query/Comment AQ \arabic{aqctr}.}}
{\par\vspace{\baselineskip}\normalcolor}

\begin{document}

\begin{frontmatter}



\title{Extending flow birefringence analysis to combined extensional--shear flows via Jeffery--Hamel flow measurements}

\author[label1]{Miu Kobayashi}
\author[label2]{William Kai Alexander Worby}
\author[label2,label3]{Misa Kawaguchi}
\author[label4]{Yuto Yokoyama}
\author[label5]{Sayaka Ichihara}
\author[label2,label6]{Yoshiyuki Tagawa\texorpdfstring{\corref{cor1}}{}}

\affiliation[label1]{organization={Department of Industrial Technology and Innovation, Tokyo University of Agriculture and Technology},
            addressline={Nakacho}, 
            city={Koganei},
            postcode={184-8588}, 
            state={Tokyo},
            country={Japan}}
\affiliation[label2]{organization={Department of Mechanical Systems Engineering, Tokyo University of Agriculture and Technology},
            addressline={Nakacho}, 
            city={Koganei},
            postcode={184-8588}, 
            state={Tokyo},
            country={Japan}}
\affiliation[label3]{organization={Institute of Engineering, Academic Assembly, Shinshu University},
            addressline={4-17-1}, 
            city={Wakasato},
            postcode={3808553}, 
            state={Nagano},
            country={Japan}}
\affiliation[label4]{organization={Micro/Bio/Nanofluidics Unit, The Okinawa Institute of Science and Technology},
            addressline={Onna-son}, 
            city={Kunigami-gun},
            postcode={904-0495}, 
            state={Okinawa},
            country={Japan}}
\affiliation[label5]{organization={Department Soft Matter, Otto-von-Guericke-University},
            city={Universitätspl. 2},
            postcode={39106}, 
            state={Magdeburg},
            country={Germany}}
\affiliation[label6]{organization={Institute of Global Innovation Research, Tokyo University of Agriculture and Technology},
            addressline={Nakacho}, 
            city={Koganei},
            postcode={184-8588}, 
            state={Tokyo},
            country={Japan}}
\cortext[cor1]{Corresponding author}
\begin{abstract}
This study investigates the relationship between phase retardation and strain rates in combined extensional–shear flows using the Jeffery–Hamel flow formalism, which yields an analytical velocity solution.  
Flow birefringence was measured in a 1.0 wt\% cellulose nanocrystal (CNC) suspension using a high-speed polarization camera.  
The velocity field was validated via particle image velocimetry (PIV), which showed good agreement with the analytical solution.  
In regions dominated by either shear or extensional components, the birefringence behavior was consistent with prior theoretical and experimental findings.  
In the combined extensional–shear regions of the Jeffery–Hamel flow, the birefringence magnitude followed the root-sum-square (RSS) of the shear- and extension-induced contributions.  
This observation aligns with the principal stress formulation derived from Mohr’s circle,  
in which the principal stress is expressed as the RSS of extensional and shear stresses.  
This finding provides a basis for extending stress–birefringence analysis to flows with coexisting deformation modes.
\end{abstract}



\begin{keyword}
Flow birefringence \sep Photoelasticity \sep Cellulose nanocrystals \sep Jeffery--Hamel flow \sep Stress-optic law \sep High-speed polarization camera
\end{keyword}

\end{frontmatter}



\section{Introduction}


The photoelastic measurement technique has emerged as an attractive method for visualizing stress distributions in fluids,  
as it enables non-intrusive, full-field stress evaluation~\citep{bawdenLiquidCrystallineSubstances1936, rankinStreamingBirefringenceStudy1989, huFlowVisualizationUsing2009, laneBirefringentPropertiesAqueous2022a,
laneTwodimensionalStrainRate2023,
nakamineFlowBirefringenceCellulose2024,  
worbyExaminationFlowBirefringence2024,Kawaguchi2025, Kusuno2025}.  
In fluids, birefringence is induced by the alignment of suspended anisotropic particles or macromolecules~\citep{peterlinOpticalEffectsFlow1976a, pihBirefringentfluidflowMethodEngineering1980, huFlowVisualizationUsing2009, 
sunMeasurementsFlowinducedBirefringence2016, hausmannDynamicsCelluloseNanocrystal2018,kadarCelluloseNanocrystalLiquid2021},  
and the resulting phase retardation can be related to the stress field through the stress-optic law (SOL)~\citep{notoApplicabilityEvaluationStressoptic2020, laneTwodimensionalStrainRate2023}. 
Previous studies have confirmed relationships between birefringence $\Delta n$ and the corresponding strain rates of the shear rate $\dot\gamma$ in simple shear flows and the extensional rate $\dot\varepsilon$ in pure extensional flows~\citep{calabreseEffectsShearingExtensional2021, laneBirefringentPropertiesAqueous2022a}.  
Recent studies have explored birefringence in more complex configurations,  
such as shear along the optical axis~\citep{worbyExaminationFlowBirefringence2024},  
the radial flow of shear-thinning fluids~\citep{Kawaguchi2025},  
and multiscale orientation dynamics in hierarchical systems under simple shear using combined rheological, polarized light, and X-ray scattering techniques.
However, in combined extensional–shear flows—which are common in practical applications—the applicability of the stress-optic law to quantify birefringence from coexisting shear and extensional deformation has not been systematically validated experimentally.

Meanwhile, although the SOL is well established in solid mechanics~\citep{sampsonStressopticLawPhotoelastic1970,prabhakaranStressopticLawOrthotropicmodel1975, rajpals.sirohiOpticalMethodsMeasurement1999,lautrePhotoelasticityApproachCharacterization2015, vivekResidualStressAnalysis2015, rameshDigitalPhotoelasticityRecent2020},
measurement techniques based on the SOL have been used to measure and visualize stress fields in soft materials~\citep{raoNewModelMaterial1955, tomlinsonPhotoelasticMaterialsMethods2015, rapetShearwaveGenerationCavitation2019, miyazakiDynamicMechanicalInteraction2021, yokoyamaIntegratedPhotoelasticitySoft2023, mitchellInvestigationHertzianContact2023}.
Nonetheless, the question of how the stress-optic law should be interpreted or extended to flows involving both shear and extensional components—such as combined extensional–shear flows—remains unexplored, particularly regarding whether principal stress formulations such as those derived from Mohr’s circle~\citep{shamesIntroductionSolidMechanics2000} can adequately describe the resulting birefringence.

This study aims to investigate the relationship between phase retardation and local strain rates in combined extensional--shear flows.  
To this end, we employ a Jeffery--Hamel flow, a two-dimensional radial flow between converging plates that naturally includes both extensional and shear components.  
Extensional flow dominates near the centerline, while shear flow dominates near the channel walls,  
allowing for spatial separation of flow types within a single geometry.
Moreover, a Jeffery--Hamel flow admits an analytical solution for the velocity field and features a simple planar geometry~\citep{jefferyTwodimensionalSteadyMotion1915, rosenheadSteadyTwodimensionalRadial1940, esmailiApproximationAnalyticalSolution2008, joneidiThreeAnalyticalMethods2010},  
making it particularly suitable for both quantitative analysis and the experimental measurement of combined extensional--shear regions.  

In this study, we conduct flow birefringence measurements using a high-speed polarization camera  
and compare the measured phase retardation distribution with the analytical velocity field of a Jeffery--Hamel flow.  
We explore whether the birefringence can be quantitatively described by extensional and shear components,  
and whether a root-sum-square formulation, consistent with principal stress theory, can describe birefringence in the combined flow regime.  
This investigation provides a fundamental step toward understanding stress–birefringence relationships in flows where multiple deformation modes coexist,  
and offers a potential framework for applying photoelastic techniques to more general, practically relevant flow environments.

\section{Theory}
\subsection{Jeffery--Hamel flow} \label{sec:J-H flow}
\begin{figure}[tb]
    \centering
    \includegraphics[width=0.3\linewidth]{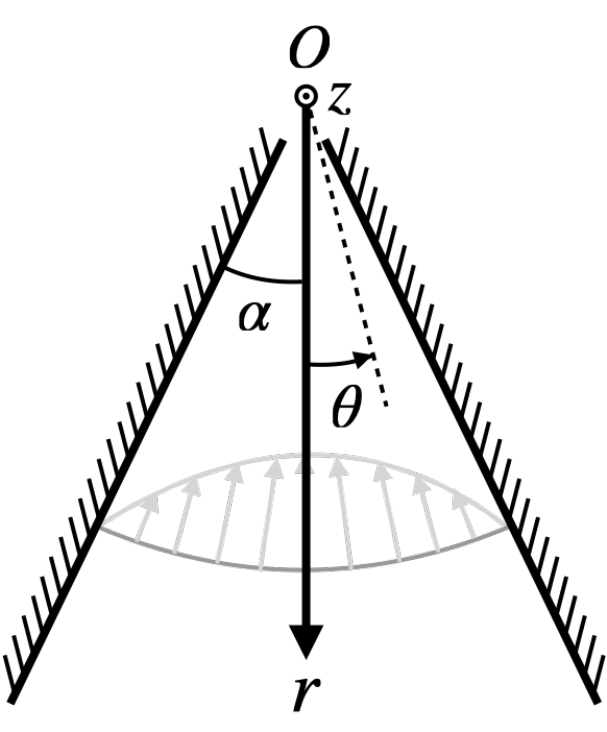}
    \caption{A Jeffery--Hamel flow in cylindrical coordinates.
    The flow is a two-dimensional radial flow with a constant plane angle ($\theta = \alpha$) and a sink or source at the origin $O$.}
    \label{fig:J-H}
\end{figure}

The Jeffery--Hamel flow is a two-dimensional flow created by two converging or diverging semi-infinite planes, with a source or sink at the point of intersection of the two plane walls~\citep{jefferyTwodimensionalSteadyMotion1915, hamelSpiralformigeBewegungenZaher1917}.
As shown in Fig.~\ref{fig:J-H}, a cylindrical coordinate system ($r$, $\theta$, $z$)  is adopted with its origin at $O$.
The two-dimensional flow is in the $r$--$\theta$ plane.
A convergent steady incompressible viscous flow is considered between two intersecting, semi-infinite planes inclined at an angle $\theta = \pm \alpha$.
The velocity components in the $r$ and $\theta$ directions are expressed as $u_r$ and $u_\theta$, respectively.
The flow is directed toward the origin $O$ and entirely radial; therefore, $u_\theta = 0$ and $u_r = u(r,\theta)$.


The continuity equation is expressed as
\begin{align}
    \frac{1}{r}\frac{\partial}{\partial r}\left( r u_r \right) = 0 ,
    \label{Eq:continuity}
\end{align}
and the momentum conservation equations are represented as
\begin{align}
    u_r\frac{\partial u_r}{\partial r}
    &=
    -\frac{1}{\rho}\frac{\partial p}{\partial r} + \nu\left( \frac{\partial^2 u_r}{\partial r^2} + \frac{1}{r}\frac{\partial u_r}{\partial r} - \frac{u_r}{r^2} + \frac{1}{r^2}\frac{\partial^2 u_r}{\partial \theta^2}\right) , \\
    0 
    &=
    -\frac{1}{\rho r}\frac{\partial p}{\partial \theta} + \frac{2\nu}{r^2}\frac{\partial u_r}{\partial \theta}.
    \label{Eq:momentum}
\end{align}

The maximum velocity $u(r, 0)$ is defined as $u_{\mathrm{max}}(r)$.
Here, defining 
$\eta = \frac{\theta}{\alpha}$,
$f(\eta)=\frac{u_r}{u_{\mathrm{max}}}$, and
$Re = u_{\mathrm{max}}\frac{r\alpha}{\nu}$,
the following third-order nonlinear differential equation is derived~\citep{whiteVISCOUSFLUIDFLOW2006}:
\begin{align}
    f''' + 2Re\alpha ff' + 4\alpha^2f' = 0 .
    \label{Eq:J-H}
\end{align}

The boundary conditions, which include a no-slip condition at the channel walls, are given as follows:
\begin{align}
    f(+1) = f(-1) = 0, \ f(0) = 1, \ f'(0) = 0 .
    \label{Eq:boundary}
\end{align}

Based on the cylindrical coordinate system $(r, \theta, z)$, and assuming $u_\theta = 0$, the velocity gradient tensor $\boldsymbol{\nabla u_r}$,  
the deformation rate tensor $\boldsymbol{S}$, and the stress tensor $\boldsymbol{\sigma}$  
are given by the following equations:
\begin{align}
    \boldsymbol{\nabla u_r} \label{eq:JH nabla}
    &=
    \renewcommand{\arraystretch}{1.7}
    \begin{bmatrix}
        \frac{\partial u_r}{\partial r} & \frac{1}{r}\frac{\partial u_r}{\partial \theta} \\
        0 & \frac{u_r}{r}
    \end{bmatrix},\\
    \boldsymbol{S} \label{eq:JH S}
    &=
    \renewcommand{\arraystretch}{1.7}
    \begin{bmatrix}
        \frac{\partial u_r}{\partial r} & \frac{1}{2r}\frac{\partial u_r}{\partial \theta} \\
        \frac{1}{2r}\frac{\partial u_r}{\partial \theta} & \frac{u_r}{r}
    \end{bmatrix},\\
    \boldsymbol{\sigma} \label{eq:JH sigma}
    &=
    -p\boldsymbol{I} + 2\mu \boldsymbol{S}\\
    &=
    \renewcommand{\arraystretch}{1.7}
    \begin{bmatrix}
        -p + 2\mu\frac{\partial u_r}{\partial r} & \mu\left(\frac{1}{r}\frac{\partial u_r}{\partial \theta}\right) \\
        \mu\left(\frac{1}{r}\frac{\partial u_r}{\partial \theta}\right) & -p + 2\mu\frac{u_r}{r}
    \end{bmatrix}\\
    &=
    -p\boldsymbol{I} + \mu 
    \begin{bmatrix}
        \dot\varepsilon & \dot\gamma \\
        \dot\gamma & -\dot\varepsilon
    \end{bmatrix}, \label{eq:JH sigma EpsilonGamma}
\end{align}
where the extensional rate $\dot\varepsilon$ and shear rate $\dot\gamma$ of the Jeffery--Hamel Flow are expressed as
\begin{align}
        \dot\varepsilon &= 2\frac{\partial u_r}{\partial r} = -2\frac{u_r}{r} \quad (\because \text{Eq.~(\ref{Eq:continuity})}), \label{Eq:epsilon}\\
        \dot\gamma &= \frac{1}{r}\frac{\partial u_r}{\partial \theta} \label{Eq:gamma}.
\end{align}

Note that this study adopts a different definition of the strain rate compared to \cite{calabreseEffectsShearingExtensional2021}.
In their work, the extensional and shear rates were defined as $\dot\varepsilon = {\partial u_x}/{\partial x}$.
In contrast, this study defines the extensional rate as $\dot\varepsilon = 2\partial u_r/\partial r$ to ensure consistency with the stress tensor formulation in Eq.~\eqref{eq:JH sigma}.
This definition allows the extensional and shear components to be expressed in a form that aligns with the symmetric structure of the strain rate tensor,
making the relationship between stress and strain more intuitive.
By adopting this definition,
the theoretical framework becomes clearer and more directly comparable to established fluid mechanics formulations.

\subsection{Theory of photoelasticity} \label{sec:Photoelasticity}
The photoelastic method is a widely used non-contact technique for evaluating stress fields, particularly in solids (see, e.g., \citet{abenIntegratedPhotoelasticityNondestructive2000}). 
It is based on the phenomenon of birefringence, which occurs in optically anisotropic materials. 
When such photoelastic materials are subjected to stress, their refractive indices vary with the polarization direction, resulting in birefringence. 
This method enables the determination of stress distributions by analyzing the phase retardation and orientation of transmitted polarized light. 
The theoretical background relating phase retardation to the stress field is provided by the SOL (\textsection\ref{sec:Stress-optic law}), while the experimental determination of phase retardation and orientation angle is described using the phase shifting method (\textsection\ref{sec:phase shifting method}). 
The fundamental physical mechanism by which stress induces birefringence is explained in Section~\ref{sec:Birefringence}.

\subsubsection{Birefringence}\label{sec:Birefringence}
Birefringence is a phenomenon in which light entering an optically anisotropic material splits into two polarized components—commonly referred to as ordinary and extraordinary rays—that travel at different speeds within the medium due to differences in refractive indices along specific directions~\citep{bornPrinciplesOpticsElectromagnetic1999}.

Certain materials, known as photoelastic materials, exhibit stress-induced birefringence.
If circularly polarized light is incident on a stress-loaded photoelastic material, birefringence introduces different phase retardations depending on the vibration direction of the light.
This results in the light becoming elliptically polarized.
Birefringence can be quantified by measuring the phase retardation ($\Delta$) and orientation angle ($\phi$) of the resulting elliptically polarized light.
$\Delta$ quantifies the optical path difference between the two orthogonal components of light, while the orientation of the principal stress directions within the material is represented by $\phi$.
These quantities are connected to the internal stress of the material through the SOL (\textsection\ref{sec:Stress-optic law}), enabling the evaluation of stress distributions.

\subsubsection{Stress-optic law (SOL)} \label{sec:Stress-optic law}

The SOL establishes the relationship between the phase retardation $\Delta$, the orientation angle $\phi$, and the stress components within a material.  
It provides a theoretical framework for connecting optically measurable quantities to the internal stress state of a material through birefringence.  
Principal stresses are characteristic values of the stress state inside a material, 
obtained through a coordinate transformation that sets
the shear stress component of the stress tensor to zero.
The difference between the maximum and minimum principal stresses, $\sigma_1$ and $\sigma_2$, is called the principal stress difference.
The phase retardation $\Delta$ corresponds to the principal stress difference,  
while the orientation angle $\phi$ is associated with the direction of the principal stress.

\begin{figure}
    \centering
    \includegraphics[width=0.8\linewidth]{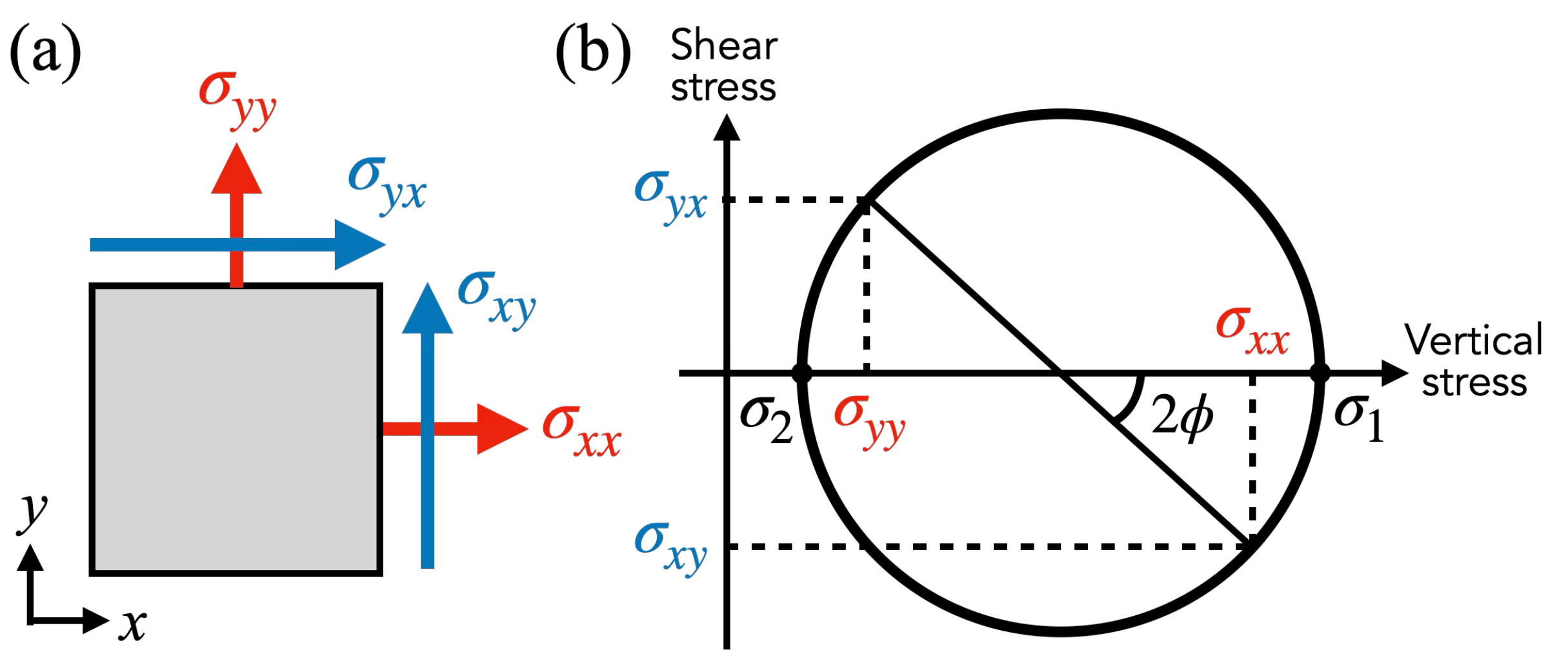}
    \caption{(a) An illustration of stresses acting in a two-dimensional plane. (b) Mohr’s circle for a two-dimensional stress system.}
    \label{fig:Mohr's circle}
\end{figure}

Using Mohr's circle, the principal stress difference and orientation angle can be determined from stress components in the Cartesian coordinate system,  
as illustrated in Fig.~\ref{fig:Mohr's circle}(a) and (b).  

The phase retardation $\Delta$ and orientation angle $\phi$ are then given by the SOL
~\citep{abenPhotoelasticityGlass1993, abenPhotoelasticityGlass2012, RameshBook}:  
\begin{align}
    \Delta &= Cd (\sigma_1 - \sigma_2) \label{Eq:sec}\\
    &= Cd\sqrt{(\sigma_{xx} - \sigma_{yy})^2 + 4\sigma_{xy}^2}\label{Eq:sol_ret}\\
    &= Cd\sqrt{(\sigma_{rr} - \sigma_{\theta\theta})^2 + 4\sigma_{r\theta}^2},\\
    \phi &= \frac{1}{2}\tan^{-1}\frac{2\sigma_{xy}}{\sigma_{xx} - \sigma_{yy}}\\
    &= \frac{1}{2}\tan^{-1}\left({\frac{(\sigma_{rr} - \sigma_{\theta\theta})\sin{2\theta} + 2\sigma_{r\theta}\cos{2\theta}}{(\sigma_{rr} - \sigma_{\theta\theta})\cos{2\theta} - 2\sigma_{r\theta}\sin{2\theta}}}\right) .
    \label{Eq:sol_phi}
\end{align}
Here, $C$ [1/Pa] is the stress-optic coefficient, $d$ [m] is the thickness of the sample along the optical axis,  
and $(\sigma_1 - \sigma_2)$ [Pa] represents the secondary principal stress difference projected onto a plane normal to the optical axis.  
These expressions are valid under the assumption that the stress distribution is uniform along the optical axis and that the principal stresses are perpendicular to it.  
Under these conditions, the birefringence $\Delta n$ can be defined as:  
\begin{align}
    \Delta n = \Delta / d .
    \label{Eq:birefringence}
\end{align}
While this formulation has been extensively validated for solid materials, its extension to fluid systems presents new challenges.

In a Jeffery–Hamel flow, the principal stress difference can be expressed in terms of the stress components and velocity gradients,  
and further reformulated by substituting in the strain rates $\dot\varepsilon$ and $\dot\gamma$ as follows:
\begin{align}
    \sigma_1 - \sigma_2
    &= \sqrt{(\sigma_{rr} - \sigma_{\theta\theta})^2 + 4\sigma_{r\theta}^2} \nonumber \\
    &= 
    \sqrt{
    \left( 2\mu\frac{\partial u_r}{\partial r} - 2\mu\frac{u_r}{r} \right)^2
    +
    4\left\{ \mu\left(\frac{1}{r}\frac{\partial u_r}{\partial \theta}\right) \right\}^2} \nonumber \\
    &= 
    2\mu\sqrt{{\dot\varepsilon}^2 + {\dot\gamma}^2}.
     \label{eq:stress}
\end{align}

If we assume that the stress-optic coefficient $C$ is constant and that the stress field arises from a combination of shear and extensional components,  
Eqs.~(\ref{eq:stress}), (\ref{Eq:sec}), and (\ref{Eq:birefringence}) lead to the following expression:
\begin{align}
    \Delta n &= C(\sigma_1 - \sigma_2) \\
    &= 2\mu C \sqrt{{\dot\varepsilon}^2 + {\dot\gamma}^2} \label{Eq:sol_Mohr}\\
    &= \sqrt{(2\mu C \dot\varepsilon)^2 + (2\mu C \dot\gamma)^2} \label{Eq:App1}\\
    &= \sqrt{\Delta n_{\dot\varepsilon}^2 + \Delta n_{\dot\gamma}^2},
    \label{Eq:sol_ret2}
\end{align}
where $\Delta n_{\dot\varepsilon} = 2\mu C \dot\varepsilon$ and $\Delta n_{\dot\gamma} = 2\mu C \dot\gamma$ denote birefringence contributions in pure extensional and pure shear flows, respectively.

This formulation suggests that, in combined extensional–shear flows, the total birefringence $\Delta n$ may be interpreted as the root-sum-square (RSS)  
of the individual contributions from extensional and shear deformation.  
This hypothesis forms the central premise of the present study and is tested experimentally in the sections that follow.

\subsubsection{Phase shifting method} \label{sec:phase shifting method}
\begin{figure}
    \centering
    \includegraphics[width=0.9\linewidth]{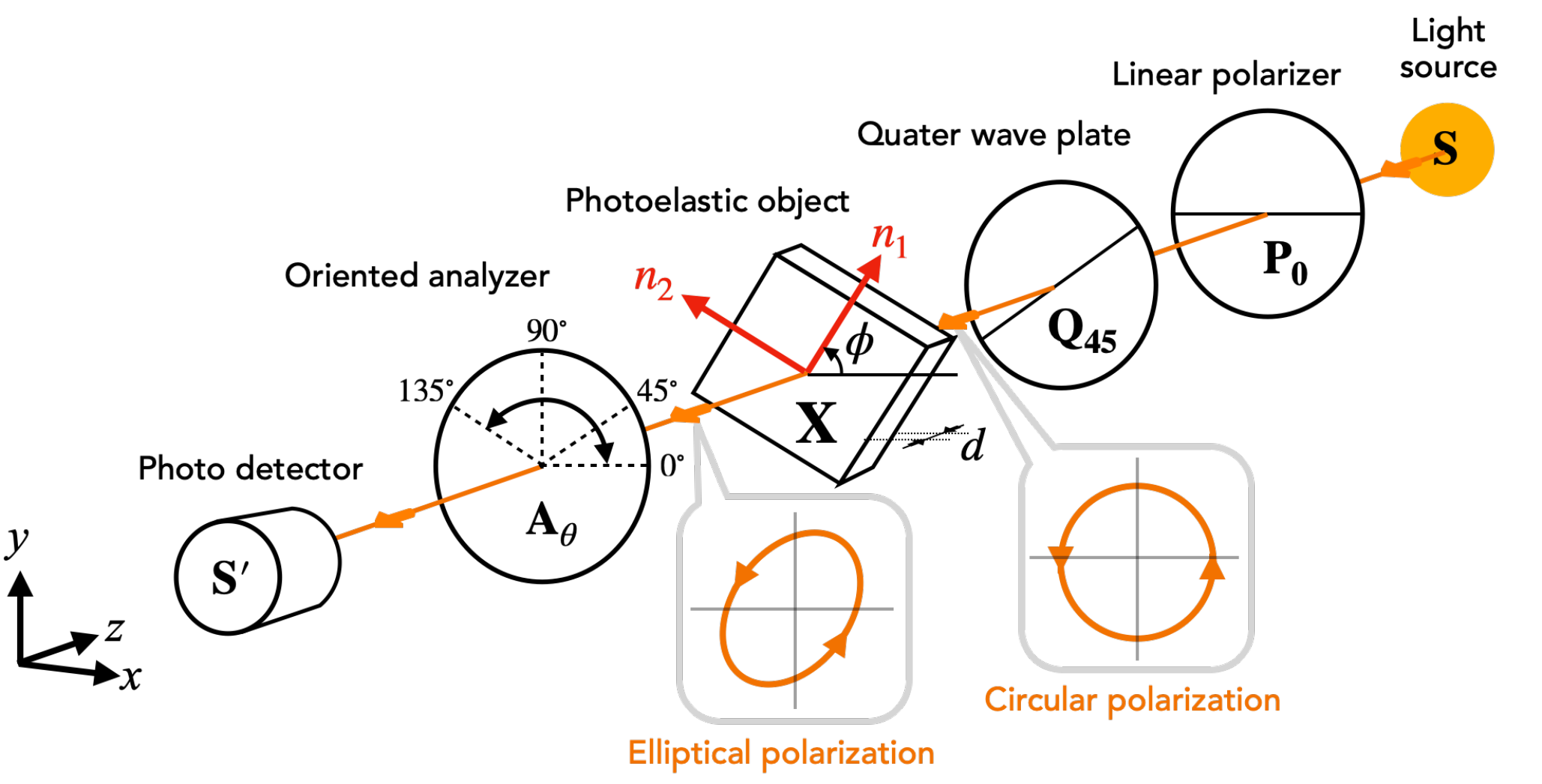}
    \caption{An illustration of the measurement system.}
    \label{fig:Photoelastic}
\end{figure}

    The measurement principle of the photoelastic method is shown in Fig.~\ref{fig:Photoelastic}.
    Circularly polarized light transforms into elliptically polarized light when transmitted through a photoelastic object experiencing birefringence.
    To measure the optical parameters ($\Delta$ and $\phi$) of elliptically polarized light, the phase shifting method~\citep{rameshDigitalPhotoelasticityRecent2020, onumaDevelopmentTwodimensionalBirefringence2014, otaniTwodimensionalBirefringenceMeasurement1994} is used.
    Light emitted from an unpolarized source first passes through a linear polarizer ($\mathbf{P_0}$) oriented at 0$^{\circ}$, followed by a quarter-wave plate ($\mathbf{Q_{45}}$) oriented at $45^\circ$, resulting in left-handed circularly polarized light.
    When the circularly polarized light enters an optically anisotropic object ($\mathbf{X_{\Delta,\phi}}$), it becomes elliptically polarized light with phase retardation $\Delta$ and orientation angle $\phi$. 
    To determine $\Delta$ and $\phi$, the rotating analyzer ($\mathbf{A_\theta}$) is adjusted to angles of $0^\circ$, $45^\circ$, $90^\circ$, and $135^\circ$, and the transmitted light intensity is measured using a photodetector. The corresponding measured light intensities are denoted as $I_0$, $I_{45}$, $I_{90}$, and $I_{135}$, respectively.
    The phase retardation and orientation angle are calculated from the measured light intensities ($I_0$, $I_{45}$, $I_{90}$, and $I_{135}$)
    ~\citep{onumaDevelopmentTwodimensionalBirefringence2014} as follows:
    \begin{align}
        \Delta &= \frac{\lambda}{2\pi}
         \sin^{-1}\frac{\sqrt{(I_{90} - I_0)^2 + (I_{45} - I_{135})^2}}{(I_0 + I_{45} + I_{90} + I_{135})/2} ,
        \label{Eq:PolarizationCamera}\\
        \phi &= \frac{1}{2}\tan^{-1}\frac{I_{90} - I_0}{I_{45} - I_{135}} ,
        \label{Eq:PolarizationCameraPhi}
    \end{align}
    where $\lambda$ is the wavelength of the light source.
    The high-speed polarization camera incorporates four polarizers oriented at angles of 
    $0^\circ$, $45^\circ$, $90^\circ$, and $135^\circ$.
    
\section{Experiments}
This section outlines the experimental methodology. 
First, an overview of the experimental setup for measuring the Jeffery--Hamel flow field, including the photoelastic measurement system, is presented in Section \ref{sec:setup}.
Section \ref{sec:PIV experiment} covers the particle image velocimetry (PIV) measurements, which were conducted to validate the two-dimensional assumption of the flow field and ensure consistency with the theoretical model.

\subsection{Experimental setup} \label{sec:setup}
\begin{figure}
    \centering
    \includegraphics[width=0.7\linewidth]{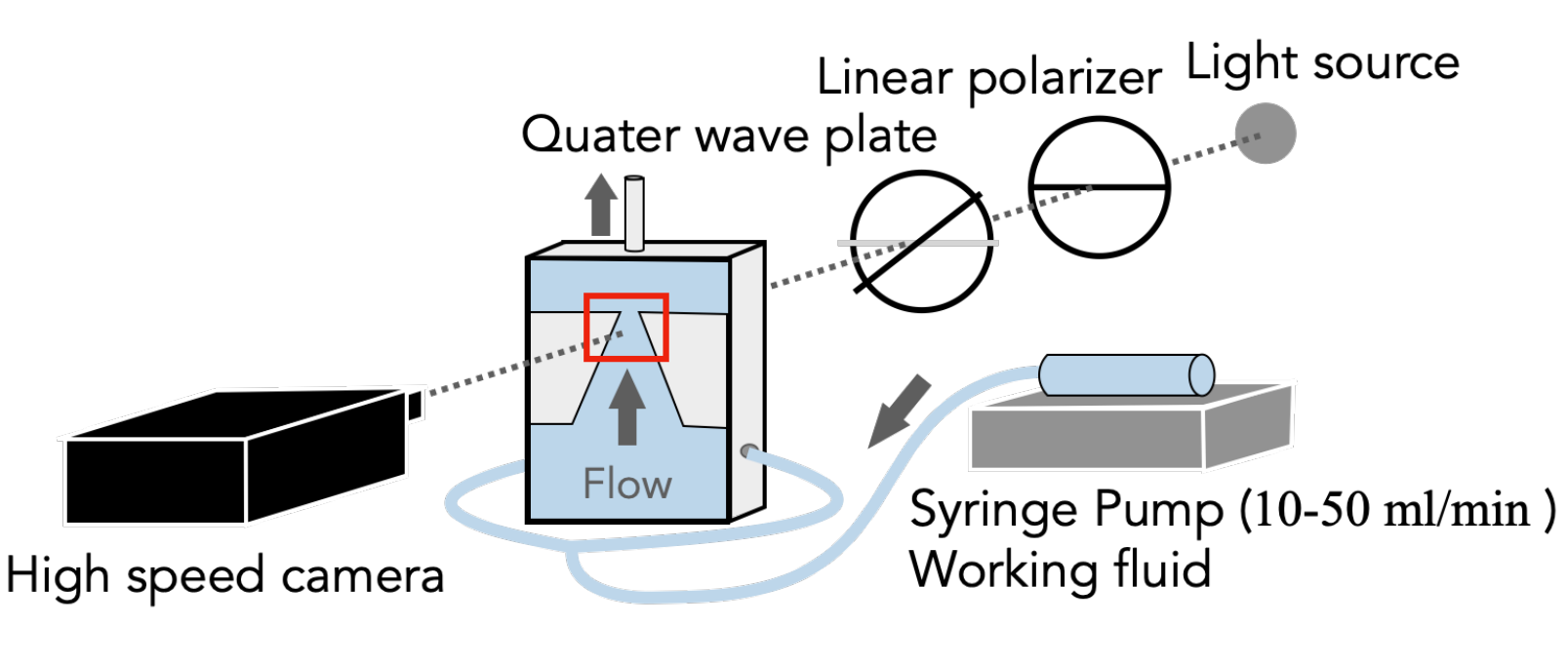}
       \centering
    \caption{A schematic of the experimental setup. The high-speed polarization camera, flow channel, and light source are aligned along the same optical axis. The channel depth is $d$ = 25 mm. The red box shows the measurement area captured by the high-speed polarized light camera during polarization measurements.
    }
    \label{fig:setup}
\end{figure}
\begin{figure}
    \centering
    \includegraphics[width=0.7\linewidth]{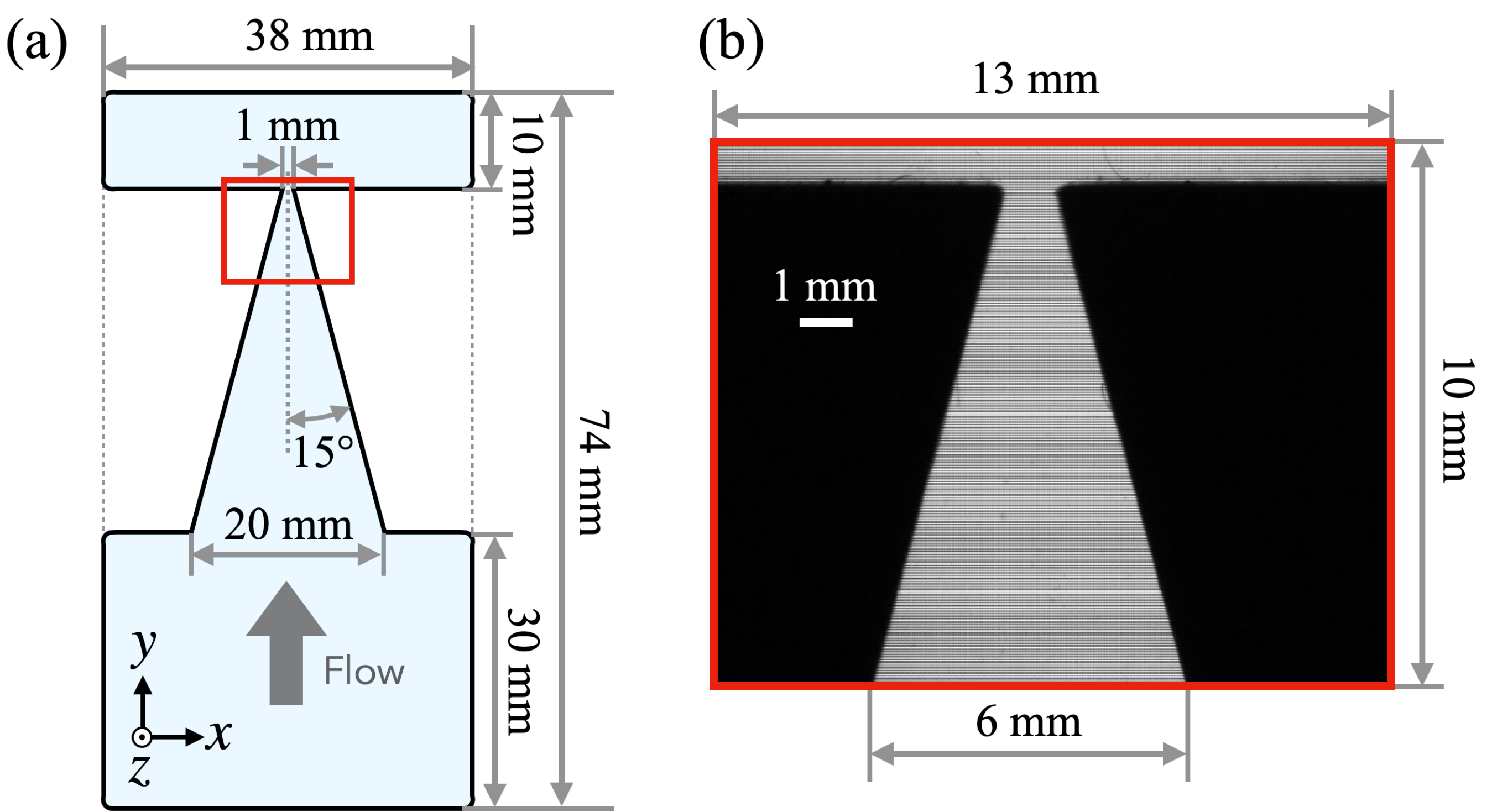}
       \centering
    \caption{(a) Geometry and key dimensions of the Jeffery--Hamel flow channel. The angle between the central axis and the converging walls is $\alpha = 15^\circ$. The width of the narrow outlet is $w$ = 1 mm, and the channel depth is $d$ = 25 mm. (b) A raw image obtained by the polarization camera with key dimensions.
    }
    \label{fig:flow channel}
\end{figure}

The experimental setup consisted of a high-speed camera, a flow channel, and a light source, all aligned along the same optical axis, as illustrated in Fig.~\ref{fig:setup}.
The flow channel included entrance and exit regions around the Jeffery--Hamel flow section.
Its geometry featured a wall angle of $\alpha = 15^\circ$, a narrowest section width of $1\ \mathrm{mm}$, and a depth of $d = 25\ \mathrm{mm}$, as shown in Fig.~\ref{fig:flow channel}.
The flow channel was fabricated using a 3D printer.
The measurement region, highlighted in red in Figs.~\ref{fig:setup} and \ref{fig:flow channel}, was sealed by two flush-mounted glass plates to ensure optical clarity. 
The flow direction was in the narrowing direction (opposite to gravity)
driven by a syringe pump (PUMP 11 ELITE, Harvard Apparatus), with flow rates of 10 and 50 ml/min, in steps of 10 ml/min.

The working fluid was a cellulose-nanocrystal (CNC) suspension (CNC-HS-FD, Cellulose Lab Co. Ltd.), consisting of rod-like nanoparticles with a width of 5--20 nm and a length of 100--250 nm.
These particles exhibit birefringence by aligning in response to stress~\citep{alizadehgiashiShearInducedAlignmentAnisotropic2018, calabreseEffectsShearingExtensional2021, laneTwodimensionalStrainRate2023}.

\begin{figure}
    \centering
    \includegraphics[width=0.6\linewidth]{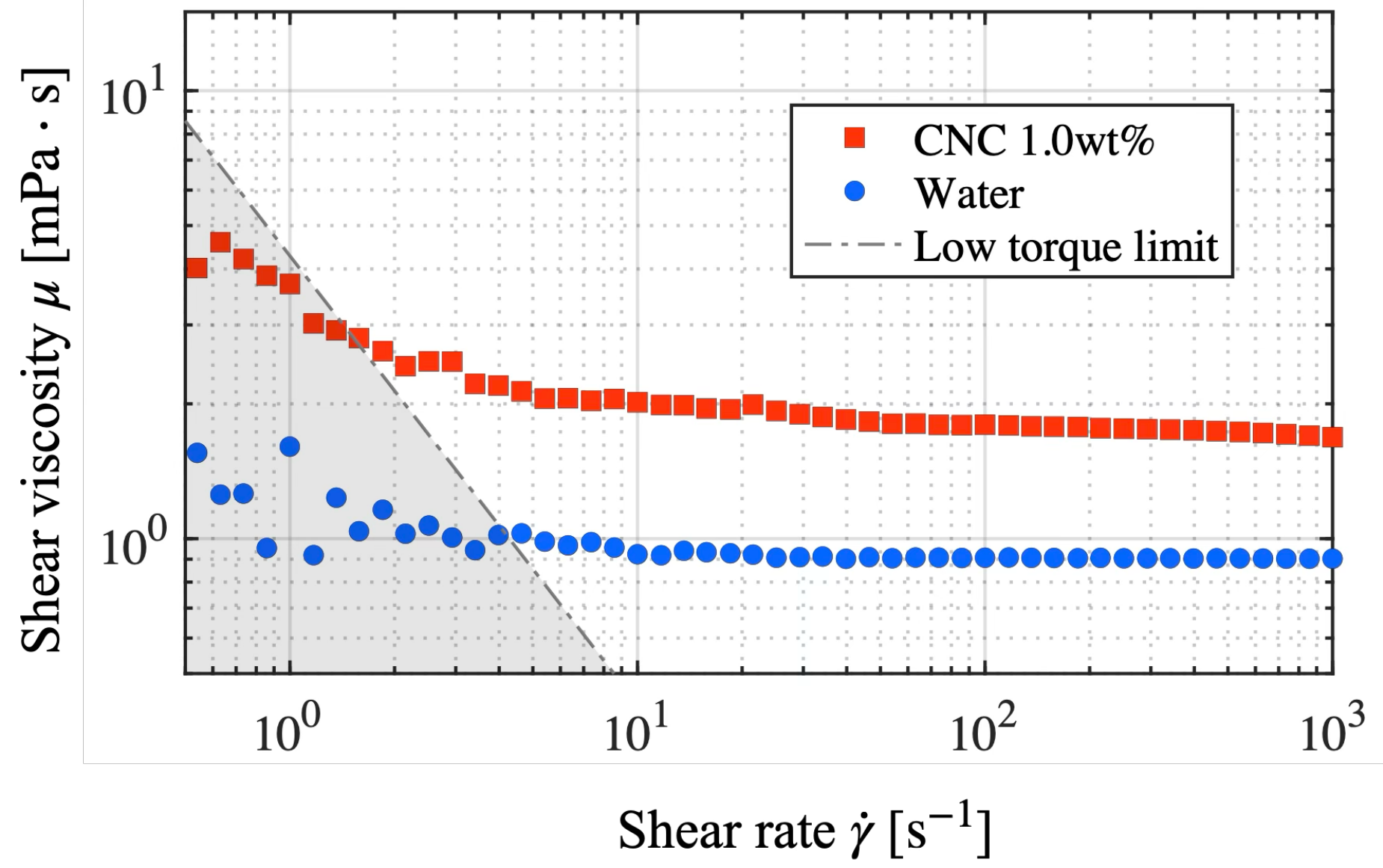}
       \centering
    \caption{The averaged shear viscosity $\mu$ from four measurements as a function of the shear rate $\dot\gamma$.
    The viscosity remains nearly constant within the measured shear rate range, indicating Newtonian fluid behavior.}
    \label{fig:ShearViscousity}
\end{figure}

The CNC suspension was prepared as follows.
Initially, CNCs were dispersed in ultra-pure water and stirred for over two hours at $25~^{\circ}\mathrm{C}$ using a stirrer (CHPS-170DF, ASONE Co., Ltd.) rotating at 600 rpm.
The concentration of the CNC suspension was set to 1.0 wt\%.
To ensure uniform dispersion of individual CNC nanorods, the CNC suspension was further sonicated for 200 seconds using a homogenizer (UX-300, Mitsui Electric Co. Ltd.).
The shear viscosity of the CNC suspension was measured using a cone-plate rheometer (MCR302, Anton Paar) at $22~^{\circ}\mathrm{C}$.  
The measurements were performed over a shear rate range from approximately $10^0$ to $10^2$ s$^{-1}$,  
and the data shown in Fig.~\ref{fig:ShearViscousity} represent the average of four individual measurements.  
As shown in Fig.~\ref{fig:ShearViscousity}, the shear viscosity remains relatively constant in the range of 1.8–2.2 mPa$\cdot$s  
for shear rates from the rheometer's low torque limit ($\sim 10^1$ s$^{-1}$) up to about $10^4$ s$^{-1}$.  
This behavior indicates that, within the range relevant to the present study, the CNC suspension can be reasonably approximated as a Newtonian fluid.
This assumption is further supported by the agreement between the experimentally measured velocity distribution and the analytically derived distribution based on Newtonian assumptions, as discussed in Section~\ref{sec:PIV experiment}.

\begin{figure}
    \centering
    \includegraphics[width=0.9\linewidth]{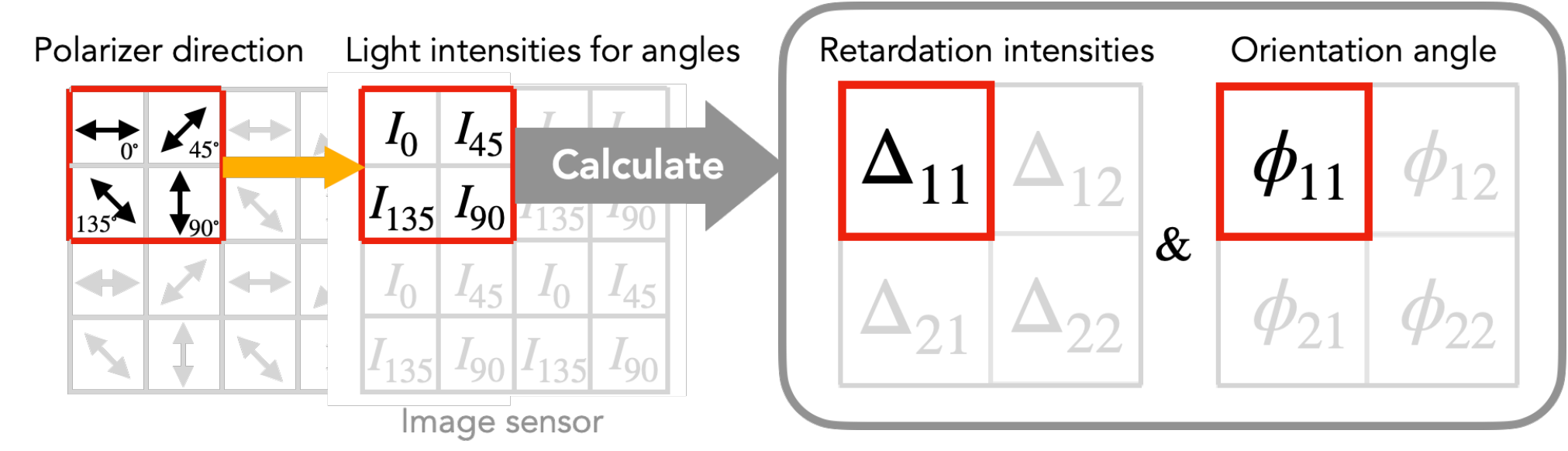}
    \caption{A schematic diagram of the measurement principle and image sensors of a high-speed polarization camera. The four polarization sensors acquire the polarization intensity and calculate the retardation $\Delta$ and the orientation angle $\phi$.
    The direction of the polarizers shown in the diagram indicates the direction of oscillation of the light transmitted.}
    \label{fig:PolarizationCamera}
\end{figure}

A high-speed polarization camera (CRYSTA PI-5WP, Photoron Co., Ltd., temporal resolution: 250 fps, spatial resolution: 848 × 680 pixels, 12-bit depth) was used to capture light-intensity images.
The flow field was recorded for two seconds, starting several tens of seconds after the solution began flowing into the flow channel to ensure steady-state conditions.
The light-intensity images obtained by the polarization camera were binarized to detect the channel walls, allowing for the determination of the flow origin in the Jeffery--Hamel flow.
A single pair of $\Delta n$ and $\phi$ values was obtained 
using the phase shifting method~\citep{rameshDigitalPhotoelasticityRecent2020, otaniTwodimensionalBirefringenceMeasurement1994} calculated by software (CRYSTA Stress Viewer, Photron Co., Ltd.)
from the brightness values of four pixels~\citep{onumaDevelopmentTwodimensionalBirefringence2014}.
Since the measurements were conducted under steady-state conditions, the time-averaged $\Delta n$ from all recorded frames was used for analysis. The experimental results were compared with the analytical solution in polar coordinates, as shown in Fig.~\ref{fig:J-H}.
    Using Eqs. (\ref{Eq:PolarizationCamera}) and (\ref{Eq:PolarizationCameraPhi}), a set of phase retardation and orientation angles was calculated from the four intensities ($I_0$, $I_{45}$, $I_{90}$, and $I_{135}$). As shown in Fig.~\ref{fig:PolarizationCamera}, the phase retardation and orientation angle derived from a 2 $\times$ 2 matrix of luminance values form a single pair, resulting in the number of phase retardation and orientation angle pixels being one-quarter of the camera's spatial resolution.
    The spatial resolution of the retardation data was 424 $\times$ 340 pixels,
    which is a quarter of 848 $\times$ 680 pixels.
    
\subsection{Validation of the velocity distribution in the channel}\label{sec:PIV experiment}
\begin{figure}
    \centering
    \includegraphics[width=0.6\linewidth]{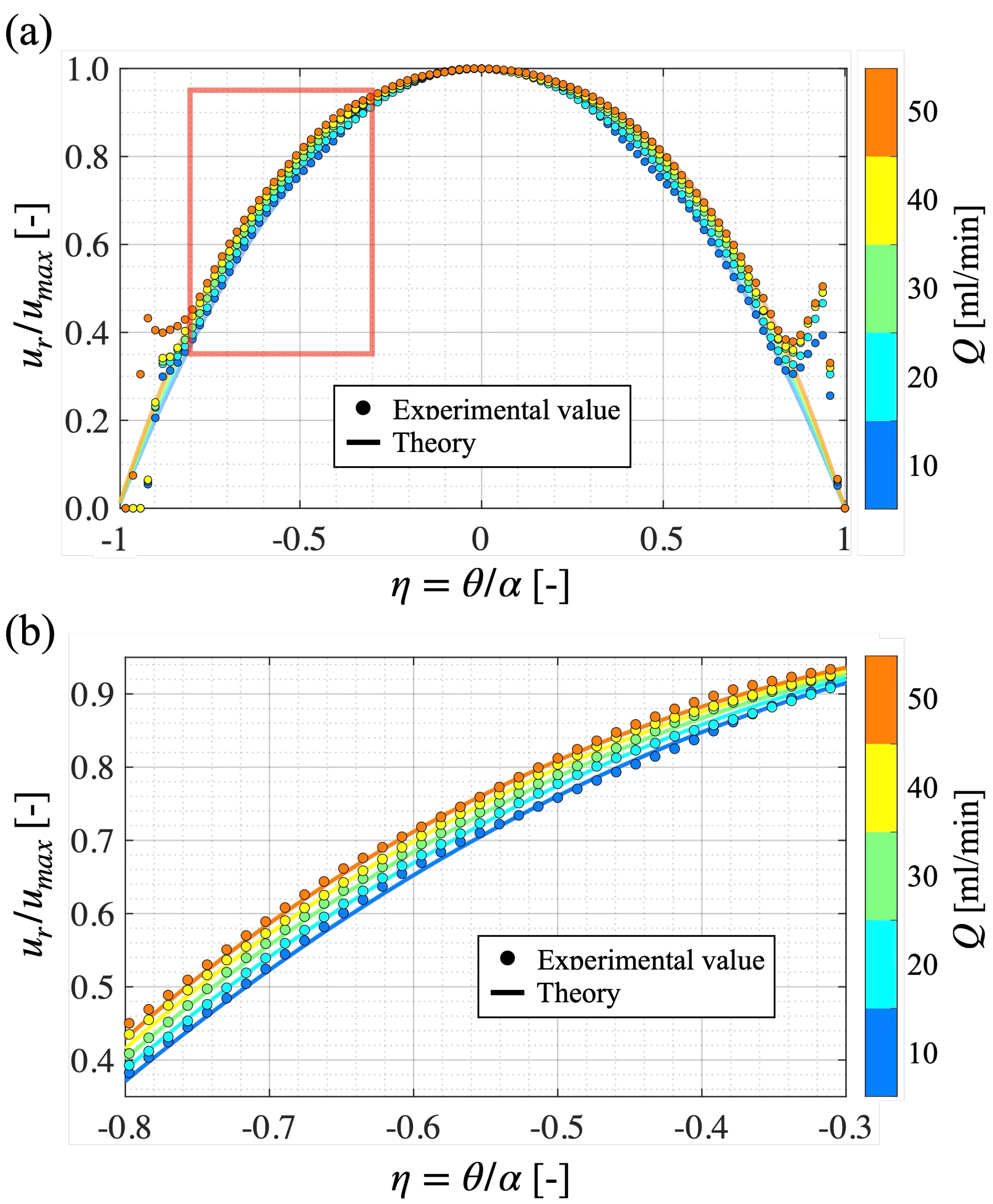}
       \centering
    \caption{(a) A comparison between the time-averaged measured velocity profiles (points) and the analytical velocity solution (solid lines) at a radius of $r$ = 3.0 mm for each flow rate $Q$; (b) Velocity profiles in the area indicated by the red box in (a).}
    \label{fig:J-H_theory}
\end{figure}


Particle image velocimetry (PIV) was performed to measure the flow field of the working fluid.  
The high-speed camera (FASTCAM SA-X2, Photron; spatial resolution: 640 × 1016 pixels), flow channel, and light source were aligned on the same optical axis.  
The camera captured shadows of tracer particles via backlighting.  
Spherical tracer particles (Toray Nylon Particulates SP-10, TORAY) with a density of $\rho = 1,080\ \mathrm{kg/m^3}$ and a diameter of $10\ \mathrm{\mu m}$ were dispersed in the 1.0 wt\% CNC suspension.  
The frame rate ranged from 125 to 1,000 fps, depending on the flow rate, to ensure that the maximum particle displacement remained within 7 pixels per frame.  

PIVlab (v3.00, Matlab) was employed to compute the velocity field using a fast Fourier transform (FFT)-based multipath cross-correlation routine, analyzing consecutive frame pairs~\citep{thielickePIVlabUserfriendlyAffordable2014, thielickeParticleImageVelocimetry2021}.
The interrogation window size was progressively refined from 64 × 64 pixels to 8 × 8 pixels, with 50\% overlap between windows to enhance spatial resolution.

A comparison between the PIV-measured velocity distribution and the analytical solution at $r = 3.0$ mm is shown in Fig.~\ref{fig:J-H_theory}.  
To facilitate comparison across different flow rates, both the experimental and analytical velocity profiles were normalized by the maximum velocity $u_{\mathrm{max}}$  
and plotted as $u_r/u_{\mathrm{max}}$.

The results demonstrate that the measured velocity distribution closely matches the analytical Jeffery--Hamel solution, with differences of less than 0.4\% in the central region.  
It is noted that velocity data near the channel walls are less accurate due to the inherent limitations of PIV in regions with steep velocity gradients and reduced particle visibility. 
Nevertheless, this does not compromise the validity of the overall comparison.  
In particular, the reasonable agreement between experimental and analytical velocity profiles in the central region suggests that the flow field was well established throughout the channel.
Furthermore, the smooth and symmetric trends observed in the birefringence measurements across the entire flow field support the assumption that deviations near the walls are minor and do not significantly affect the analysis.

The overall agreement between the Newtonian-based analytical solution and experiment remains sufficiently close to consider the flow field well characterized for the purpose of subsequent analysis.
Thus, the extensional rate $\dot\varepsilon$ and the shear rate $\dot\gamma$ were calculated from the analytical solution of the velocity field (Eqs.~(\ref{Eq:epsilon}) and~(\ref{Eq:gamma})).

\section{Results and discussion}

This section presents a detailed analysis of the flow-induced birefringence observed in a Jeffery--Hamel flow.  
We begin by examining the overall spatial distribution of the birefringence and orientation angle across various flow rates (\textsection\ref{sec:FlowRateBirefringence}) to
establish the reproducibility and reliability of the polarization measurements.  
This analysis serves as a foundation for the subsequent quantitative evaluations in shear-, extension-, and combined-flow regions.
We then validate the strain rate dependence of birefringence in the limiting cases of pure extensional and simple shear flows (\textsection\ref{sec:ShearExtensionalRegions}), confirming consistency with theoretical predictions and previous studies.  
Building on these findings, we finally focus on the combined shear--extensional region (\textsection\ref{sec:CombinedRegion}), where we reveal a quantitative relationship that describes the birefringence as the RSS of the contributions from the shear and extensional components.  


\subsection{Flow-rate dependence of the birefringence field}
\label{sec:FlowRateBirefringence}
\begin{figure*}
    \centering
    \includegraphics[width=1\linewidth]{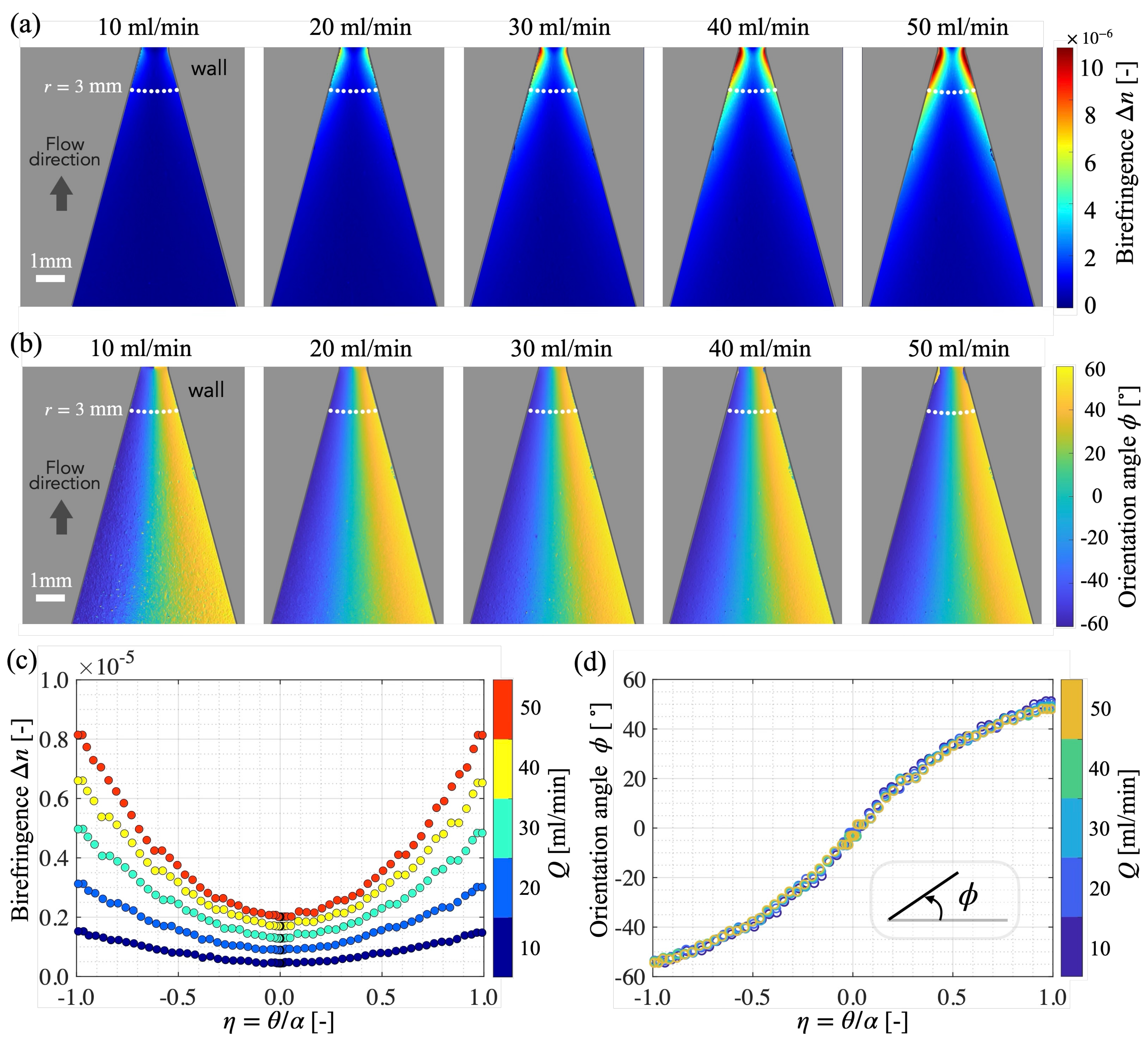}
       \centering
    \caption{(a) Distributions of the time-averaged flow birefringence $\Delta n$ and (b) orientation angle $\phi$ obtained from polarization measurements at various flow rates $Q$ (10--50 ml/min). (c) Line profiles of $\Delta n$ and (d) $\phi$ along each arc at $r$ = 3.0 mm, which are shown as white dotted lines in (a) and (b).}
    \label{fig:ResultsDistribution}
\end{figure*}

We first examine how the spatial distribution of the birefringence $\Delta n$ and orientation angle $\phi$ varies with flow rate in a Jeffery–Hamel flow.  
This provides a visual basis for understanding the flow--optical response relationship.
Figure~\ref{fig:ResultsDistribution} presents polarization measurement results for five different flow rates ($Q = 10$–50 ml/min).  
Panels (a) and (b) show the time-averaged two-dimensional distributions of $\Delta n$ and $\phi$, respectively,  
while panels (c) and (d) display the corresponding one-dimensional profiles along an arc at $r = 3.0$ mm, as indicated by the white dotted lines.

Across all flow rates, the birefringence $\Delta n$ exhibits a smooth and symmetric pattern with respect to the centerline [Fig.~\ref{fig:ResultsDistribution}(a)].  
As $Q$ increases, the overall magnitude of $\Delta n$ rises, particularly near the channel walls where higher shear rates are expected.  
These trends are clearly observed in the line profiles along the angular coordinate $\eta = \theta/\alpha$ [Fig.~\ref{fig:ResultsDistribution}(c)].  
In all cases, $\Delta n$ increases monotonically from the center ($\eta = 0$) to the wall ($|\eta| = 1$),  
indicating a consistent spatial gradient in the local strain rate and particle alignment.  
The separation between each profile confirms the strong $Q$ dependence of the birefringence amplitude.  
Because the subsequent sections analyze the dependence of $\Delta n$ on local strain rates in detail,  
this clear and structured distribution provides a strong foundation for interpreting those results.

The orientation angle $\phi$ also shows a smooth radial gradient from approximately $0^\circ$ at the channel center to around $60^\circ$ near the wall [Fig.~\ref{fig:ResultsDistribution}(b, d)],  
as expected from the direction of the principal stress.  
The consistency across different flow rates supports the steady and well-defined nature of the underlying flow field.

Overall, the results in Fig.~\ref{fig:ResultsDistribution} demonstrate that $\Delta n$ exhibits reproducible and well-structured spatial distributions,  
confirming the reliability of the polarization measurements and providing a basis for subsequent analysis in shear-, extension-, and combined-flow regimes.

\subsection{Birefringence in extensional- and shear-dominant regions}
\label{sec:ShearExtensionalRegions}
\begin{figure}
    \centering
    \includegraphics[width=0.7\linewidth]{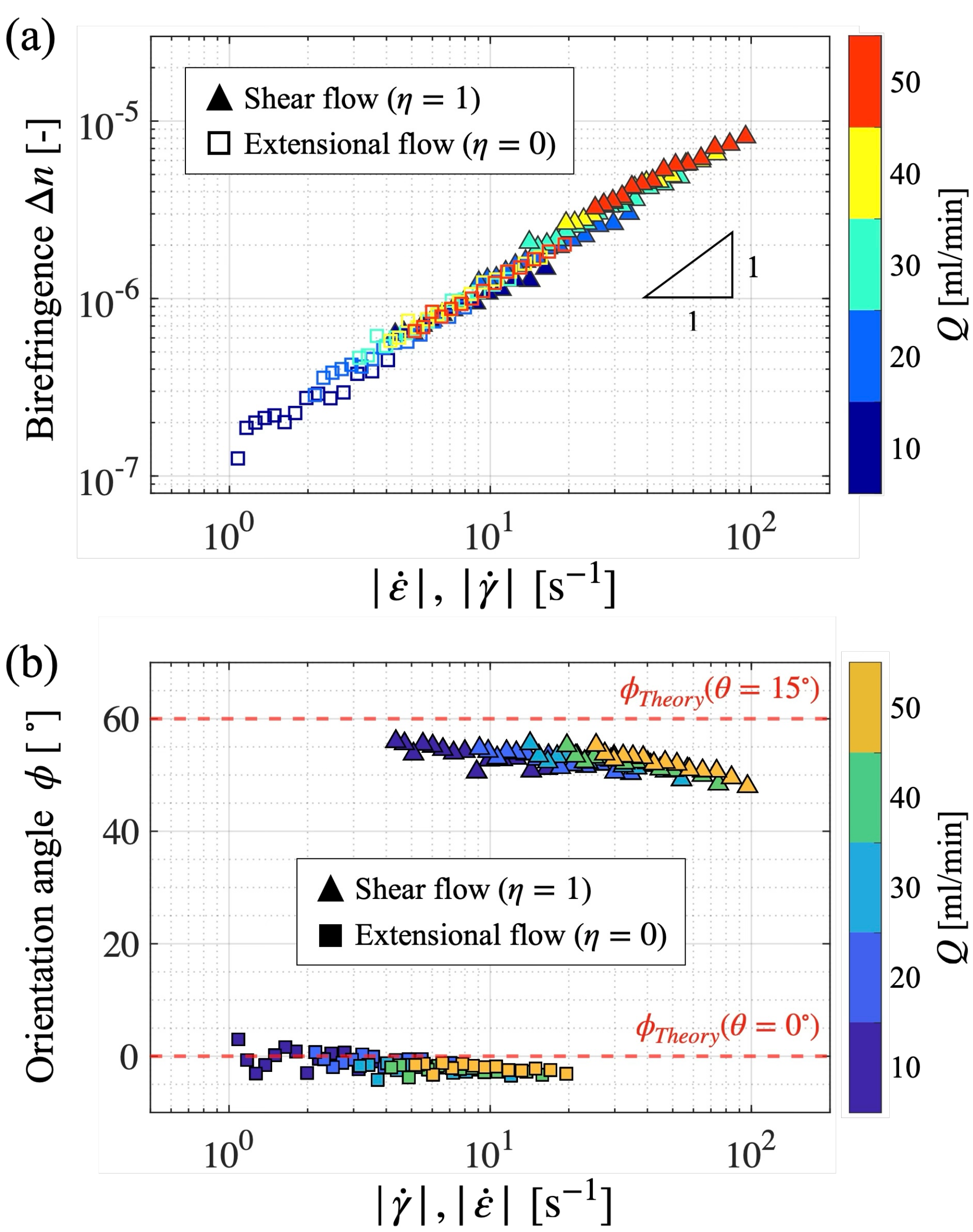}
       \centering
    \caption{(a) The time-averaged birefringence $\Delta n$, and (b) orientation angle $\phi$, as a function of the magnitude of the shear rate $|\dot\gamma|$ or extensional rate $|\dot\varepsilon|$.}
    \label{fig:shear-extensional}
\end{figure}
Before investigating the combined shear--extensional region,  
we confirmed that the birefringence $\Delta n$ behaves as expected in the limiting cases of pure extensional and simple shear flows.
To this end, we focused on regions along the channel centerline ($\eta = 0$) and near the wall ($\eta = 1$),  
where either the extensional or shear rate dominated, respectively.  
The relationship between $\Delta n$ and the local strain rates $\dot\varepsilon$ or $\dot\gamma$ was evaluated using data at $r = 3.0$–$6.0$ mm.  
As shown in Fig.~\ref{fig:shear-extensional}(a), the data collapsed onto master curves in both regions.  

Fitting the results showed 
that $\Delta n$ scales approximately linearly with strain rate in both flow types.  
This observation is consistent with the behavior predicted by Eq.~(\ref{Eq:sol_Mohr}) under conditions where either extensional or shear rate dominates,  
which simplifies to $\Delta n = 2\mu C |\dot\varepsilon|$ or $\Delta n = 2\mu C |\dot\gamma|$, respectively.  
Fitting the experimental data to this linear form yields a stress-optic coefficient of $C = (3.0 \pm 0.4)\times 10^{-5}\ \mathrm{mPa\cdot s}$.
These results are in close agreement with previous studies using CNC suspensions under similar conditions.  
Notably, while the present study assumes a linear relationship between birefringence and strain rate, previous studies reported slightly lower power-law exponents, typically around 0.9~\citep{calabreseEffectsShearingExtensional2021}.  
Such variations are reasonably attributed to differences in particle shape, concentration, or measurement geometry~\citep{laneBirefringentPropertiesAqueous2022a, worbyExaminationFlowBirefringence2024}.

The orientation angle $\phi$ also exhibited behavior consistent with theoretical expectations.  
In the extensional-dominant region ($\eta = 0$), $\phi$ was close to $0^\circ$,
indicating alignment of the anisotropic particles along the flow direction, as predicted by Jeffery’s theory
~\citep{jeffery1922motion} and confirmed in prior experiments
~\citep{calabreseEffectsShearingExtensional2021}. 
In the shear-dominant region ($\eta = 1$), $\phi$ was observed to range from $47^\circ$ to $55^\circ$,  
which is slightly below the theoretical value $\phi_{\mathrm{Theory}} = 60^\circ$ derived from the principal stress direction (Eq.~(\ref{Eq:sol_phi})).
The slight underestimation of $\phi$ compared to the theoretical prediction is consistent with earlier studies,  
which reported that as the shear strain rate increases, particle alignment tends to deviate from the ideal principal stress direction  
due to enhanced particle alignment
~\citep{Vermant2001RheoopticalDetermination, laneBirefringentPropertiesAqueous2022a}.  

Taken together, these results verify that the birefringence and orientation data in the extensional- and shear-dominant regions are consistent with both theoretical predictions and prior experimental observations.  
This establishes the reliability of our measurement approach and forms a sound basis for analyzing the more complex combined shear--extensional region discussed in the next section.

\subsection{Birefringence of the combined extensional--shear region}
\label{sec:CombinedRegion}
\begin{figure*}
    \centering
    \includegraphics[width=1\linewidth]{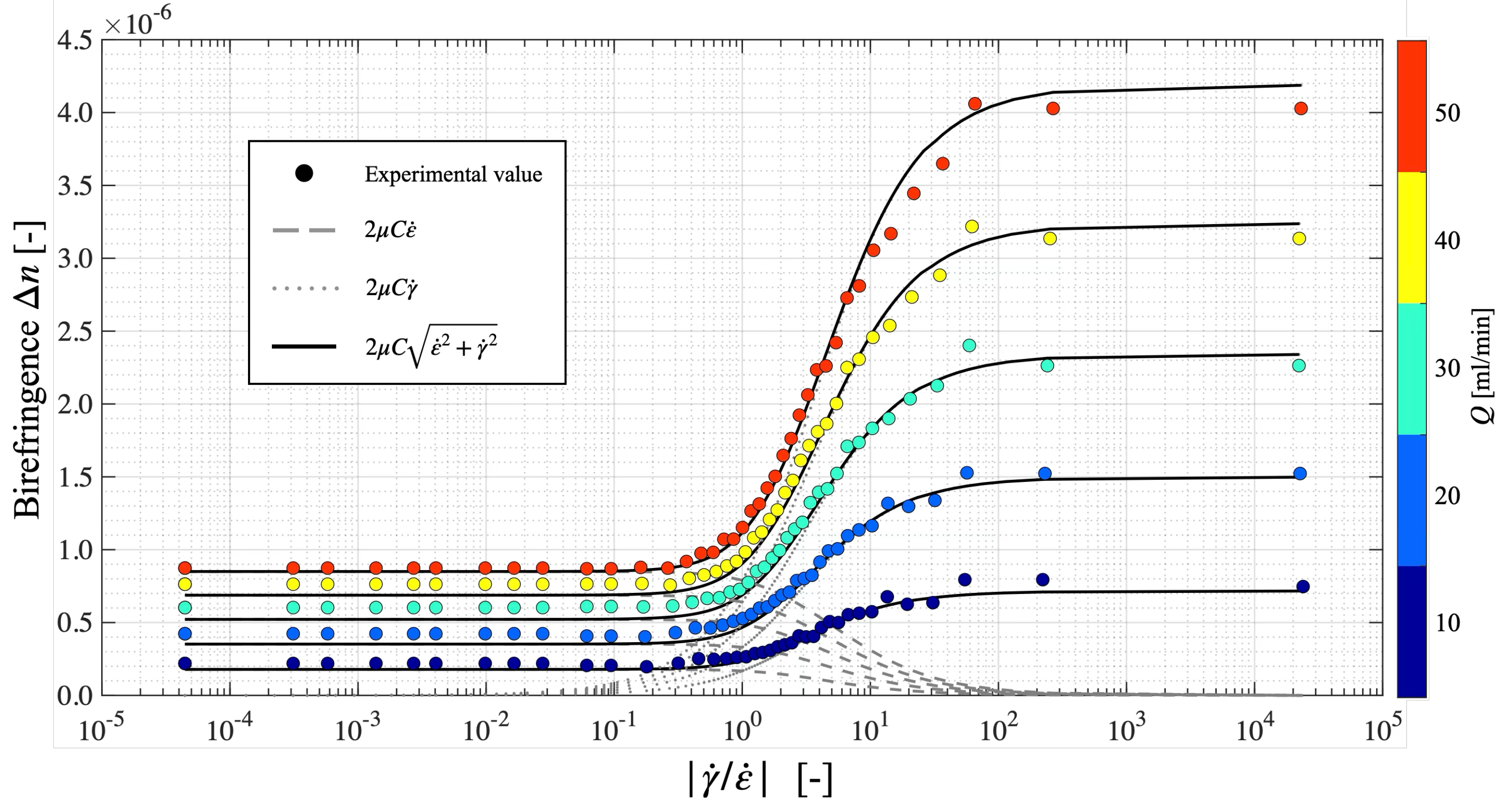}
       \centering
    \caption{The relationship between the measured birefringence $\Delta n$ and the ratio of the shear and extensional rate $\dot\gamma/\dot\varepsilon$ obtained from the analytical solution.
    The plots are experimental results and the colors correspond to flow rates at $r = 5.0\ \mathrm{mm}$.
    The dashed line represents a relation with respect to $\dot\varepsilon$,
    and the dotted line corresponds to a linear relation with $\dot\gamma$.
    The solid black line shows the root-sum-square (RSS) model with parameters $C = 3.0 \times 10^{-5}\ \mathrm{Pa^{-1}}$ and $\mu = 2.0\ \mathrm{mPa\cdot s}$.}
    \label{fig:all}
\end{figure*}

Figure~\ref{fig:all} plots the measured birefringence $\Delta n$ against the strain-rate ratio $|\dot\gamma/\dot\varepsilon|$, evaluated along an arc at $r = 5.0\ \mathrm{mm}$.  
Smaller values of this ratio correspond to extensional-dominant regions,  
while larger values indicate shear-dominant behavior.  
The intermediate range around $|\dot\gamma/\dot\varepsilon| \approx 1$ defines the combined shear–extensional region of interest.
In this region, $\Delta n$ exhibits a smooth transition between the two limiting behaviors observed in pure shear and extensional flows.  
This sigmoidal trend is consistently observed across all tested flow rates ($Q = 10$–50 ml/min), indicating that the combined effect of extensional and shear deformation manifests itself as a continuous and reproducible modulation in birefringence.
Notably, this trend is well captured by the root-sum-square (RSS) model derived from the stress-optic law (Eq.~(\ref{Eq:sol_Mohr})),  
shown as the solid black curve in Fig.~\ref{fig:all}.  
The model not only reproduces the asymptotic behavior in the limit cases but also accurately captures the intermediate regime.  
This agreement strongly suggests that the combined birefringence response can be interpreted as the geometric sum of individual shear- and extension-induced contributions.

To evaluate the predictive capability of the RSS model, we first fitted the limiting regions—five points each from the shear- and extension-dominant zones—using  
$\Delta n = 2\mu C |\dot{\gamma}|$ and $\Delta n = 2\mu C |\dot{\varepsilon}|$.  
With the known viscosity $\mu = 2.0\ \mathrm{mPa\cdot s}$, we obtained a fitted value of $C = 3.0 \times 10^{-5}\ \mathrm{Pa^{-1}}$.  
These fitted trends (dashed lines in Fig.~\ref{fig:all}) match the corresponding data within an average deviation of 10.7\% in the limiting regions.

In contrast, in the combined region, the RSS expression matches the experimental data with an even smaller deviation—just 6.7\% on average.  
This improved agreement in the most complex part of the flow further validates the RSS-based formulation  
as a robust and physically meaningful approach to describing birefringence in flows involving coexisting deformation modes.

To evaluate the robustness of this relationship, we extended the analysis to other radial positions at $r = 6.0$, $4.0$, and $3.0\ \mathrm{mm}$.  
For each position, the same procedure was applied: the coefficient $C$ was first obtained by fitting the limiting regions (shear- and extension-dominant),  
and then the resulting RSS expression was used to predict the birefringence in the combined region.

\begin{figure}
    \centering
    \includegraphics[width=0.7\linewidth]{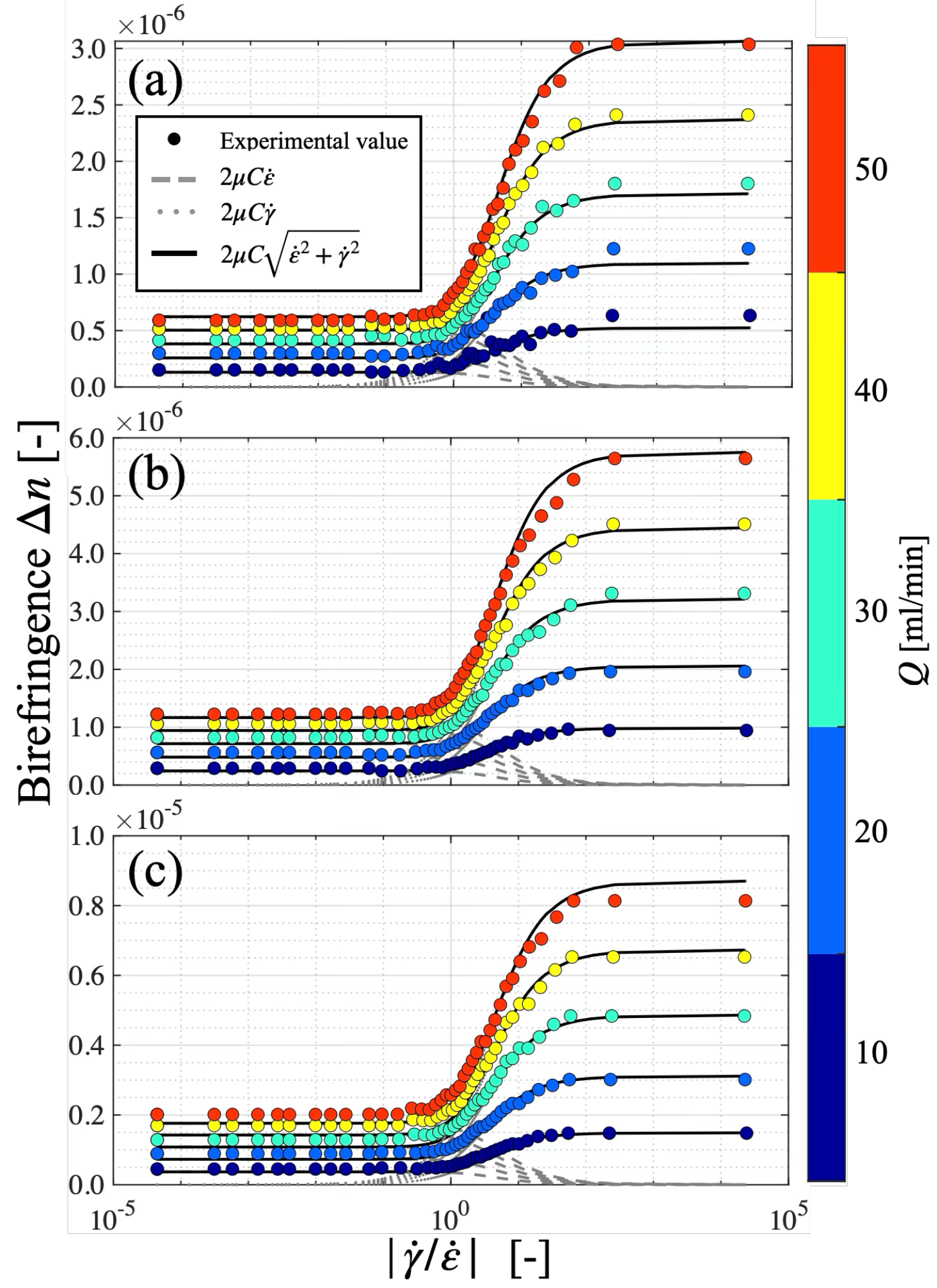}
       \centering
    \caption{The relationship between the measured birefringence $\Delta n$ and the ratio of the shear and extensional rate $\dot\gamma/\dot\varepsilon$ obtained from an analytical solution. Each panel corresponds to a different radial position: (a) $r = 6.0~\mathrm{mm}$; (b) $r = 4.0~\mathrm{mm}$; (c) $r = 3.0~\mathrm{mm}$. The solid line represents the root-sum-square  (RSS) of the two equations, which the measured birefringence consistently matches in the intermediate region plotted using $ \mu = 2.0\ \mathrm{mPa\cdot s}$ and $C = 3.2 \times 10^{-5}$, $2.6 \times 10^{-5}$, and $2.4 \times 10^{-5}\ \mathrm{Pa^{-1}}$ for (a), (b), and (c), respectively.}
    \label{fig:r456}
\end{figure}

As shown in Fig.~\ref{fig:r456}, the RSS-based formulation (Eq.~(\ref{Eq:sol_Mohr}))  
provides good agreement with the experimental data across all radial positions.  
As in the previous analysis, the deviation from the fitted curve was used to evaluate the agreement.  
The average deviation in the shear- and extensional-dominant regions was 9.6\%, whereas in the combined region it was lower at 6.5\%,  
demonstrating that the RSS-based model not only remains valid at different cross-sections,  
but captures the birefringence behavior even better in regions where both strain components coexist.

The fitted values of $C$ were found to vary slightly across positions—$3.2 \times 10^{-5}$, $2.6 \times 10^{-5}$, and $2.4 \times 10^{-5}\ \mathrm{Pa^{-1}}$ for $r=6.0$, 4.0, and 3.0 mm, respectively.  
This variation is considered acceptable and does not detract from the main conclusion,  
given that the experimental behavior was only approximately linear, while the fitting assumed a strictly linear dependence in the limiting regions.

To further explore the validity of the RSS formulation,  
we applied the trial fitting approach described in~\ref{app1},  
in which we relaxed the assumption of linear scaling in the dominant regimes.  
Instead of assuming $\Delta n \propto \dot\varepsilon$ and $\Delta n \propto \dot\gamma$ with a fixed exponent of 1,  
we allowed the exponents and coefficients in each regime to be determined via power-law fitting.  
These more flexible expressions were then combined using the same RSS form of Eq. (\ref{Eq:sol_ret2}).
The results show that, compared to the original 6.5\% deviation obtained using fixed exponents (i.e., assuming linear scaling), employing best-fit power laws in the shear- and extensional-dominant regions reduces the deviation in the combined region to 4.3\% (see~\ref{app1} for details).
This finding reinforces the idea that the RSS-based formulation is not only empirically consistent,  
but also robust across a range of fitting assumptions and strain-rate conditions.

In summary, the RSS-based formulation provides a consistent and physically interpretable description of flow-induced birefringence in combined shear–extensional flows.  
This conclusion is supported by both experimental evidence across multiple radial positions and its theoretical grounding in the stress-optic law and Mohr’s circle formalism,  
which have been well established in solid-state photoelasticity.  
The close agreement between the RSS model and experimental data, particularly in the combined region,  
suggests that this approach may serve as a general framework for analyzing birefringence in flow fields involving coexisting deformation modes.  
These findings represent an important step toward developing a unified understanding of stress-optical behavior in complex fluid flows.  
Further validation under different geometries, fluid types, and flow conditions will be essential to fully assess the generality and limitations of this framework.


\section{Conclusion}
This study has aimed to clarify the relationship between photoelastic measurements and velocity fields in combined extensional--shear flows. 
To achieve this, we conducted polarization measurements in a steady Jeffery--Hamel flow, which features both extensional and shear components and admits an analytical velocity solution.
The phase retardation $\Delta$ was measured using a high-speed polarization camera and converted to birefringence $\Delta n$. The velocity field was numerically calculated and validated via particle image velocimetry (PIV).
In extensional-dominant and shear-dominant regions, $\Delta n$ scales approximately linearly with the strain rate in both flow types.
These results are consistent with previous studies on the simple shear and pure extensional flow configurations.

Importantly, in the combined extensional--shear region, the measured birefringence was well captured by the root-sum-square expression $2\mu C\sqrt{{\dot\varepsilon}^2 + {\dot\gamma}^2}$.
This formulation is consistent with the principal stress derived from Mohr’s circle, suggesting that the birefringence in combined flows reflects the magnitude of the principal stress composed of both shear and extensional components.
These findings demonstrate, for the first time, a physically interpretable and quantitatively consistent framework for flow-induced birefringence in combined extensional--shear fields. 
This work may help extend the applicability of stress-optic methods to more complex flow environments where traditional constitutive modeling remains challenging.

\appendix
\section{A trial investigation of birefringence under a combined extensional--shear flow}
\label{app1}

To derive a more detailed empirical relationship for birefringence under combined extensional--shear flows,  
we begin the analysis using data obtained at $r = 5.0\ \mathrm{mm}$.  

\begin{figure*}
    \centering
    \includegraphics[width=0.9\linewidth]{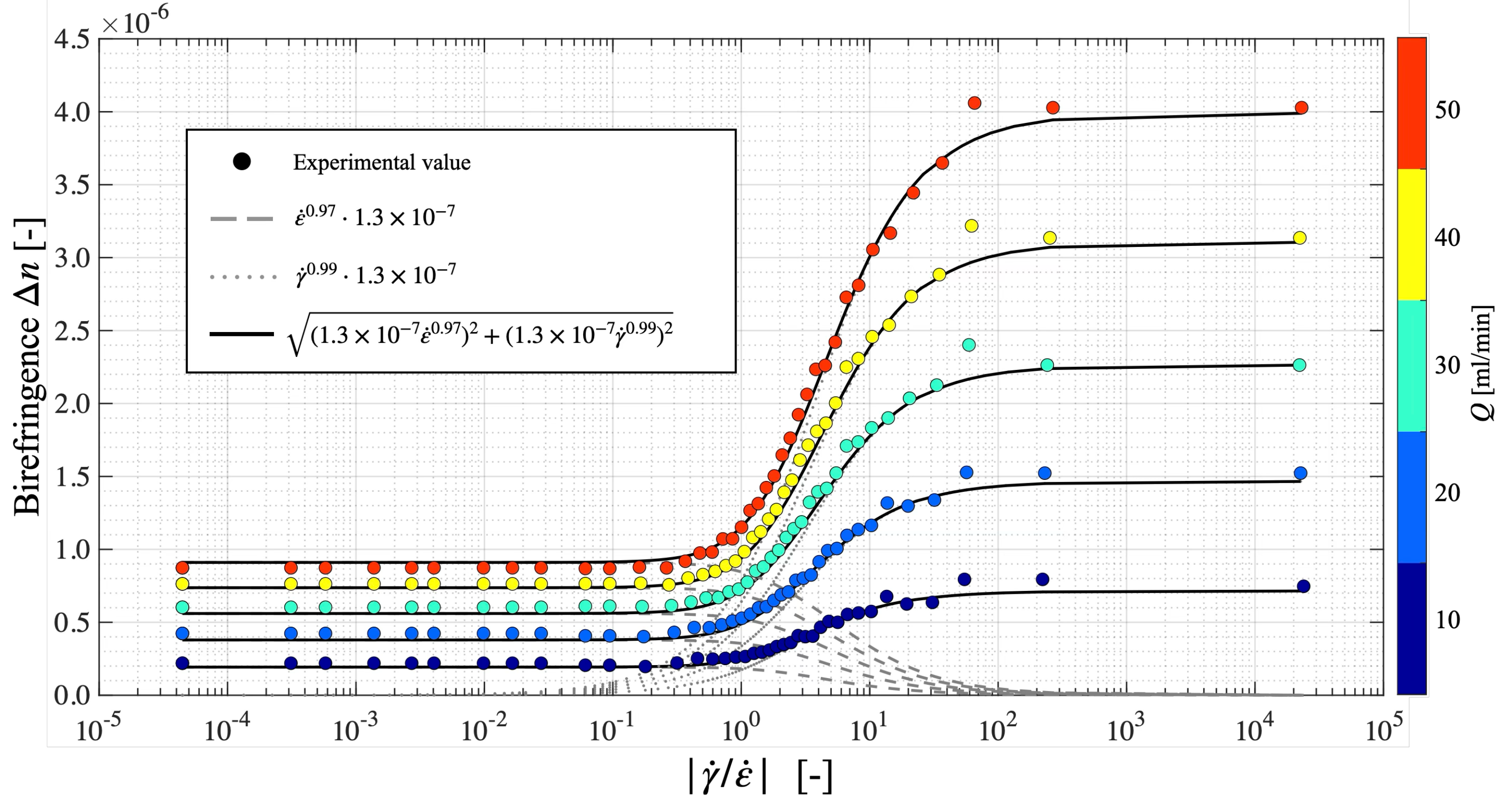}
       \centering
    \caption{The relationship between measured birefringence $\Delta n$ and the ratio of shear and extensional rate $\dot\gamma/\dot\varepsilon$ obtained from the analytical solution at $r = 5.0\ \mathrm{mm}$.
    The plots show the experimental results, and the colors correspond to flow rates.
    The dashed line shows the power law trends of $\Delta n \propto \dot\varepsilon^{0.99}$,
    and the dotted line shows the power law trend of $\Delta n \propto \dot\gamma^{0.99}$.
    The solid line is the root-sum-square of the two equations.}
    \label{fig:Appendix_all}
\end{figure*}

To characterize the limiting behaviors,  
we selected five data points at the lowest and highest $|\dot\gamma/\dot\varepsilon|$ values and performed power-law fitting.  
The extensional-dominant region yielded a relationship of $\Delta n_{\dot\varepsilon} = \dot\varepsilon^{0.97} \cdot 1.3 \times 10^{-7}$, shown as a dotted line in Fig.~\ref{fig:Appendix_all}, while the shear-dominant region gave $\Delta n_{\dot\gamma} = \dot\gamma^{0.99} \cdot 1.3 \times 10^{-7}$, shown as a dashed line.  
In both regimes—$|\dot\gamma/\dot\varepsilon| < 10^{-1}$ and $|\dot\gamma/\dot\varepsilon| > 10^{1}$—the experimental data follow the respective power-law trends well.
The average deviation in the dominant region was 5.7\%, which is lower than the 10.7\% deviation obtained in the main text under the assumption of linear scaling, indicating improved agreement with the experimental data.

The main interest lies in the intermediate region, where both shear and extensional strains coexist.  
Here, we examine whether the birefringence $\Delta n$ can be described as a combined effect of both strain components.  
According to the SOL discussed in Sec.~\ref{sec:Stress-optic law},  
the principal stress difference in a Jeffery--Hamel flow may be approximated by a root-sum-square (RSS) of the shear and extensional contributions.  
By substituting the fitted expressions for $\Delta n_{\dot\varepsilon}$ and $\Delta n_{\dot\gamma}$ into this form,  
we obtain the following empirical expression for $\Delta n$ in the combined-flow region:

\begin{align}
    \Delta n 
             &= \sqrt{(\dot\varepsilon^{0.99} \cdot 1.3 \times 10^{-7})^2 + (\dot\gamma^{0.99} \cdot 1.3 \times 10^{-7})^2}.
    \label{eq:combined_rss}
\end{align}
This expression is represented by the solid line in Fig.~\ref{fig:Appendix_all}.  
The experimental results show excellent agreement with this RSS-based formulation across the entire range of $|\dot\gamma/\dot\varepsilon|$,  
including the transition region between the shear- and extension-dominant regions.  

\begin{figure}
    \centering
    \includegraphics[width=0.7\linewidth]{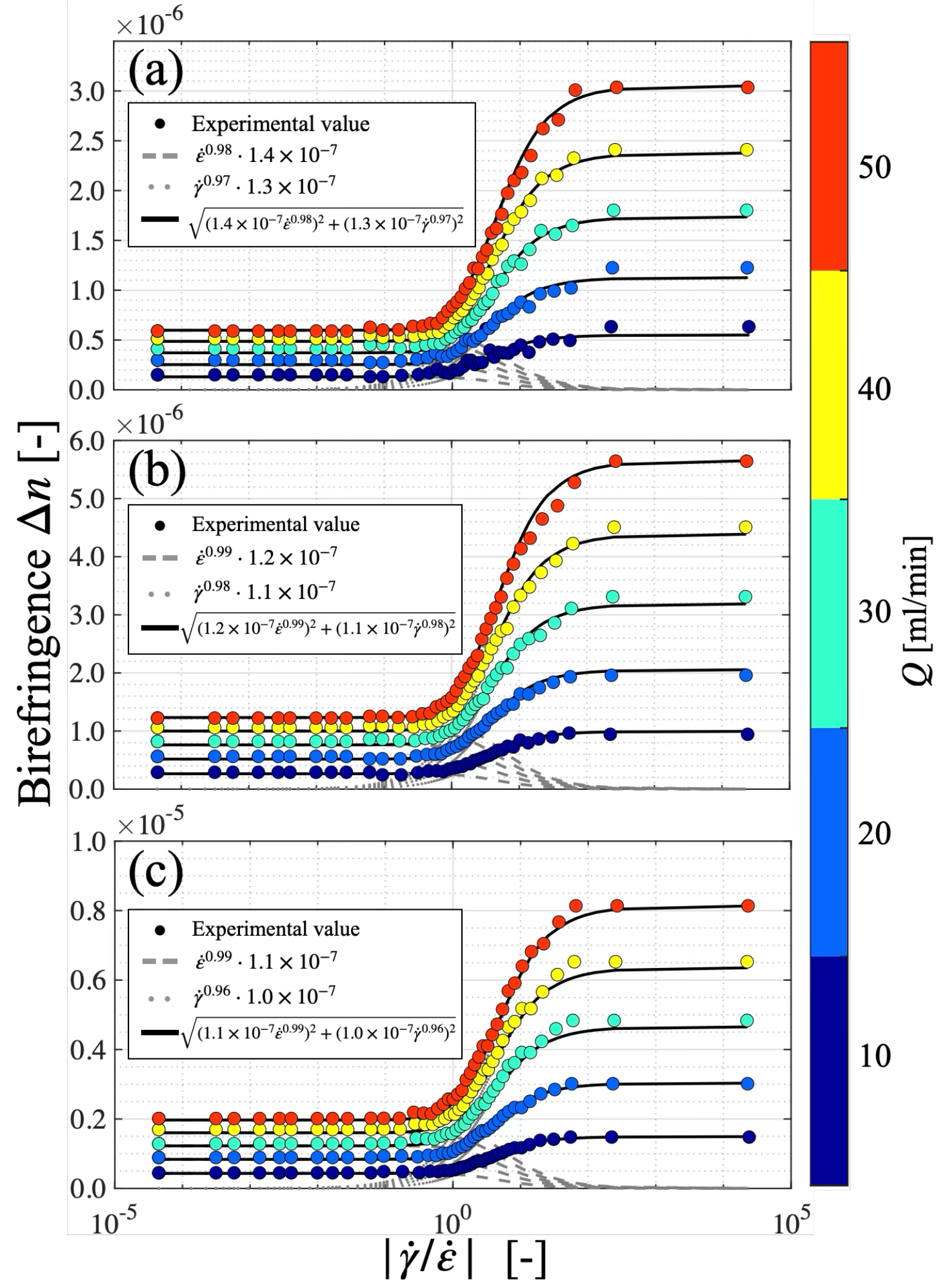}
       \centering
    \caption{The relationship between the measured birefringence $\Delta n$ and the ratio of the shear and extensional rates $\dot\gamma/\dot\varepsilon$ obtained from the analytical solution. The solid line represents the root-sum-square of the two equations, which the measured birefringence consistently matches in the intermediate region. Each panel corresponds to a different radial position: (a) $r = 6.0 \mathrm{mm}$; (b) $r = 4.0 \mathrm{mm}$; (c) $r = 3.0 \mathrm{mm}$.}
    \label{fig:Appendix_r456}
\end{figure}

To evaluate the robustness of this empirical relationship, we extended the analysis to additional radial positions at $r = 6.0$, $4.0$, and $3.0\ \mathrm{mm}$.  
For each position, the same procedure was applied to extract $\Delta n_{\dot\varepsilon}$ and $\Delta n_{\dot\gamma}$ in the extensional and shear-dominant regions,  
followed by the construction of the RSS expression for the combined region.  
As shown in Fig.~\ref{fig:Appendix_r456}, in all cases, the measured birefringence in the intermediate region was well described by the expression  
$\Delta n = \sqrt{ \Delta n_{\dot\varepsilon}^2 + \Delta n_{\dot\gamma}^2 }$ (Eq.~(\ref{Eq:sol_ret2})),  
confirming that the same RSS-based formulation applies beyond a single cross-section.
In the combined region, the average deviation was further reduced to 4.3\%, demonstrating even better agreement than in the dominant regions.

Therefore, we conclude that the RSS-based formulation constructed from the limiting behaviors of $\Delta n_{\dot\varepsilon}$ and $\Delta n_{\dot\gamma}$  
offers a consistent and interpretable description of birefringence in combined extensional--shear flows, at least within the Jeffery--Hamel flow field considered here.


\section*{CRediT authorship contribution statement}
Miu Kobayashi: Investigation, Experiments, Visualization, Methodology, Interpretation, Writing – original draft, review and editing.
William Kai Alexander Worby: Investigation, Methodology, Interpretation, Writing – review and editing.
Misa Kawaguchi: Investigation, Methodology, Interpretation, Writing – review and editing.
Yuto Yokoyama: Investigation, Methodology, Interpretation, Writing – review and editing.
Sayaka Ichihara: Interpretation, Writing – review and editing.
Yoshiyuki Tagawa: Conceptualization, Methodology, Funding acquisition, Supervision, Writing – review and editing.

All authors agree to be accountable for all aspects of the work in ensuring that questions related to the accuracy or integrity of any part of the work are appropriately investigated and resolved.



\section*{Acknowledgement}
This work was supported by the Japan Society for the Promotion of Science (JSPS) KAKENHI (Grant Nos.
JP20H00222, JP20H00223, JP22KJ1239, JP24H00289); the Japan Science and Technology Agency (JST) PRESTO (Grant No. JPMJPR2105); and the Japan Agency for Medical Research and Development (AMED) (Grant No. JP22he0422016).
We thank Dr. C. Lane for his valuable suggestions on the flow channel design and for providing the flow channel.

 \bibliographystyle{elsarticle-num-names} 
 \bibliography{bibliography}

\end{document}